\documentclass[journal]{vgtc}                     

\onlineid{0}

\vgtccategory{Research}

\vgtcpapertype{please specify}

\title{Viewpoint-Tolerant Depth Perception for Shared Extended Space Experience on Wall-Sized Display}

\author{%
  \authororcid{Dooyoung Kim}{0000-0002-6003-2181}, 
  \authororcid{Jinseok Hong}{0009-0008-4647-5016}, 
  \authororcid{Heejeong Ko}{0000-0001-6898-4104}, and \authororcid{Woontack Woo}{0000-0002-5501-4421}
}

\authorfooter{
  \item
  	Dooyoung Kim is with KAIST KI-ITC ARRC.
  	E-mail: dooyoung.kim@kaist.ac.kr.
  \item
  	Jinseok Hong is with KAIST UVR Lab.
  	E-mail: jindogliani@kaist.ac.kr.
  \item
  	Heejeong Ko is with KAIST UVR Lab.
  	E-mail: hj.ko@kaist.ac.kr.

  \item Woontack Woo is with KAIST UVR Lab and KAIST KI-ITC ARRC.
  	E-mail: wwoo@kaist.ac.kr.
}

\abstract{%
  We proposed viewpoint-tolerant shared depth perception without individual tracking by leveraging human cognitive compensation in universally 3D rendered images on a wall-sized display. While traditional 3D-perception-enabled display systems have primarily focused on single-user scenarios—adapting rendering based on head and eye tracking—the use of wall-sized displays to extend spatial experiences and support perceptually coherent multi-user interactions remains underexplored. We investigated the effects of virtual depths ($d_v$) and absolute viewing distance ($d_a$) on human cognitive compensation factors (perceived distance difference, viewing angle threshold, and perceived presence) to construct the wall display-based eXtended Reality (XR) space. Results show that participants experienced a compelling depth perception even from off-center angles of 23°–37°, and largely increasing virtual depth worsens depth perception and presence factors, highlighting the importance of balancing extended depth of virtual space and viewing distance from the wall-sized display. Drawing on these findings, wall-sized displays in venues such as museums, galleries, and classrooms can evolve beyond 2D information sharing to offer immersive, spatially extended group experiences without individualized tracking or wearables.
}


\teaser{
  \centering
  \includegraphics[width=\linewidth]{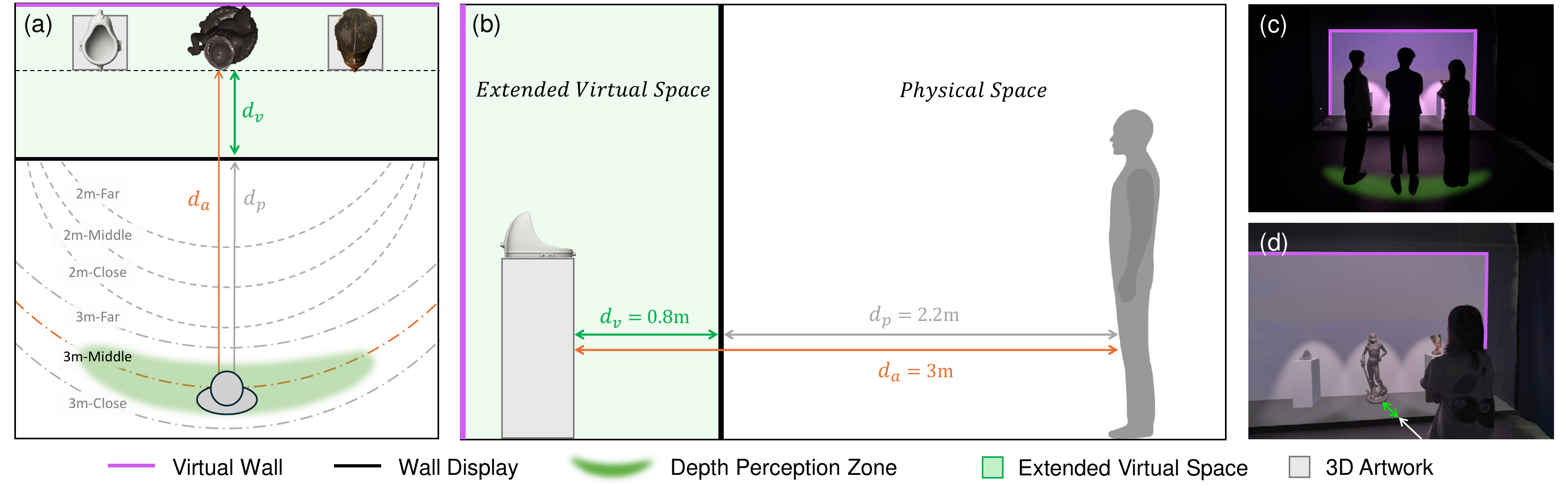}
  \caption{(a) Top view and (b) side view of a wall display-based XR space. Human cognitive ability enables perceiving the image on a large wall display as an extended virtual space (a, c) within an estimated depth perception zone (b, d), with the appropriate ratio of a virtual depth ($d_{v}$) to a physical distance ($d_{p}$) at a given absolute viewing distance ($d_{a}$).}
  \label{fig:teaser}
}

\graphicspath{{figs/}{figures/}{pictures/}{images/}{./}} 

\usepackage{tabu}                      
\usepackage{booktabs}                  
\usepackage{lipsum}                    
\usepackage{mwe}                       

\usepackage{mathptmx}                  

\usepackage{enumitem}
\usepackage{cite}
\usepackage{amsmath}
\usepackage{multirow}
\usepackage{verbatim}

\begin{document}


\firstsection{Introduction}

\maketitle

As immersive technologies continue to blur the boundaries between physical and virtual environments, wall-sized displays have gained attention for their ability to support multi-user engagement and extend spatial experiences beyond the limitations of head-mounted displays (HMDs). In venues like museums, public installations, and collaborative workspaces, these large static screens provide a shared canvas onto which expanded virtual content can be seamlessly integrated with the physical surroundings. Previous research on wall displays has often focused on information scalability—displaying large-scale data or visually partitioned content for multiple viewers—rather than delivering a fully three-dimensional, immersive experience~\cite{james2023evaluating, Cavallo2019DataspaceAR}. However, as demand grows for collective eXtended Reality (XR) experiences that preserve depth cues and spatial coherence, wall-sized displays are increasingly positioned to offer group-level immersion with users’ natural viewing perspectives~\cite{james2023evaluating, Cavallo2019DataspaceAR}. This potential has sparked renewed interest in harnessing wall displays as spatial extension media, particularly in settings like exhibitions, museums, and co-located collaborative environments. Yet, current depth solutions for such displays often rely on single-user tracking and adaptive rendering~\cite{cruz2023surround}, a constraint that inherently limits multi-user scalability and leads to suboptimal experiences for off-center viewers.

Despite the growing prominence of wall-sized displays for shared immersive experiences, creating a compelling depth perception for multiple viewers without head-mounted devices remains a significant design challenge. Traditional head-coupled image approaches through large displays typically rely on single-user head or eye tracking to ensure correct binocular cues~\cite{lawrence2024project, shen2023virtual}, thereby constraining their utility for group-based or public installations. Even multi-user solutions often require additional devices—such as stereoscopic glasses or personalized shutter systems—to deliver individual perspectives, complicating deployment in large-scale settings~\cite{pollock2012right}. While prior studies have examined the efficacy of pictorial cues and perspective convergence to enhance depth perception in static images~\cite{saunders2006accuracy, sweet2011depth}, many of these investigations were conducted under controlled, single-viewer conditions. Consequently, the question of how well depth integrity is maintained when observers with varied viewing angles share the same display has remained underexplored. To address this gap, our work investigates the tolerance of human perception to a fixed-viewpoint rendering. We examine how manipulating virtual depth and viewing distance can preserve a shared sense of depth across a range of vantage points, relying on cognitive compensation rather than individualized rendering.

In this study, we examine the perceptual feasibility of using a single, fixed-viewpoint 3D image rendered from a universal setting (eye height = $1.5\,\mathrm{m}$) to create a shared XR experience through a wall-sized display. \cref{fig:teaser} shows our wall display-based XR space. We conducted a within-subject user study using a Trompe-l'œil-style image~\cite{gombrich1960art} that rendered a 3D scene from a fixed viewpoint on a large rear-projected wall screen ($4.2\,\mathrm{m}$ × $2.36\,\mathrm{m}$). We created six experimental conditions by combining the absolute viewing distance ($d_a$: $2\,\mathrm{m}$, $3\,\mathrm{m}$) and the virtual depth of objects in the extended space ($d_v$: Close = $0.4\,\mathrm{m}$, Middle = $0.8\,\mathrm{m}$, Far = $1.2\,\mathrm{m}$). Participants assessed three perceptual metrics: \textit{perceived distance difference (PDD)}, \textit{viewpoint angular threshold (VAT)}, and perceived presence. By doing so, we leverage the innate capacity of human depth perception to tolerate deviations from an optimal viewing distance and angle, thereby distinguishing the convincing depth perception zone for multi-user access to extended virtual spaces. 

Our findings from the evaluation support this approach: participants reported acceptable depth perception even when viewing the display from off-center angles between 23° and 37°, indicating that the convincing depth perception region is broader than the traditionally assumed sweet spot. Notably, many described the loss of depth perception beyond this range as abrupt rather than gradual, suggesting sharply defined perceptual boundaries. These results demonstrate the feasibility of enabling shared immersive experiences without adaptive rendering or individualized tracking, simply by aligning display parameters with the natural tolerance of human vision. In line with previous research, we also found that PDD was increasingly underestimated as $d_v$ increased, while $d_a$ had no significant effect on PDD. Moreover, perceived presence factors—such as realness, spatial presence, and visibility—tended to worsen with increasing $d_v$. These experimental findings highlight the importance of aligning display configurations with human perceptual limits to support effective shared immersive experiences. While increasing $d_v$ can help convey a larger spatial extent, excessively large values may impair depth perception, underscoring the need to carefully balance viewer distance and perceived depth. \cref{fig:teaser}(a,b) conceptually illustrates that under our \textit{3\,m}-\textit{Middle} experimental condition, it is possible to achieve a sufficient off-center angle for viewing while maintaining a high level of perceived presence.

The contributions of this study can be threefold. First, we provide empirical evidence that a 3D scene rendered from a fixed viewpoint can support shared depth perception across a wider range of viewing angles than previously assumed, demonstrating the potential of perceptual tolerance as a foundation for wearable-free multi-user XR. Second, we present a quantitative analysis of how absolute viewing distance ($d_a$) and virtual depth ($d_v$) interact to influence depth perception, viewing angle threshold, and the user’s sense of perceived presence, offering insights into how visual parameters affect immersive experience design. Lastly, we analyze these insights into practical design implications for constructing XR environments using wall displays, enabling more accessible, inclusive, and immersive multi-user experiences in public and architectural spaces without relying on head-mounted or adaptive systems.

\section{Related Works}

\subsection{Depth Perception}

Human depth perception relies on binocular disparity along with multiple monocular cues (\textit{e.g.}, perspective, size, occlusion)~\cite{howard2002depth, gregory2015eye}. In an ideal stereoscopic display, each user's eyes receive two images corresponding to the correct viewpoint, yielding a compelling 3D illusion. While geometric models predict noticeable distortions when viewers deviate from the intended “center of projection,” studies show that people often perceive only mild depth distortion even when viewing from off-center or on mismatched screen sizes~\cite{allison2015perceptual}. Additionally, depth perception can be achieved through monocular pictorial cues, such as vanishing points, shadows, perspective, and color contrast. These cues alone can create a sufficient sense of depth, even when viewed binocularly~\cite{ichikawa2003integration}. This perceptual robustness in depth perception has been attributed to cognitive correction mechanisms, which operate without relying solely on geometric precision. Vishwanath et al.~\cite{vishwanath2005pictures} proposed a scene-inference model in which viewers reconstruct 3D layouts by combining image-based depth cues with internalized expectations, rather than relying solely on geometric accuracy. These findings suggest that higher-order perceptual processes support a stable sense of 3D structure under spatial distortions—such as off-axis viewing or non-standard screen sizes—by compensating for geometric discrepancies~\cite{cutting1986shape,cutting1987rigidity}.

In XR environments, researchers have explored how visual cues enhance depth perception. Heinrich et al.~\cite{heinrich2019depth} analyzed the effects of shadows and contrast in AR, while Hornsey et al.~\cite{hornsey2021contributions} examined effective pictorial cues in VR. Vertical disparity tolerance is known to be extremely limited—only a few arcminutes—before fusion breaks~\cite{jin2006tolerance}, and although horizontal tolerance is greater, depth can still appear distorted beyond certain angles. These findings suggest that within a certain viewing range, multiple observers may still perceive consistent 3D structure from a fixed perspective. Our study directly probes this perceptual tolerance window for viewpoint variation, building on the premise that humans can share a 3D experience without personalized rendering.

\subsection{Immersive Displays for Depth Perception}

Prior display research has largely relied on optics-based depth rendering techniques derived from the Plenoptic Function~\cite{bergen1991plenoptic}. Systems such as head-mounted displays (HMDs), and light-field displays provide real-time depth perception from multiple viewpoints. More recently, multi-user directional backlight panels have enabled glasses-free 3D viewing for multiple simultaneous viewers without resolution loss~\cite{li2022multi}. While such systems support real-time depth perception, they rely on complex optics, head tracking, and dedicated hardware, resulting in high installation costs and limited flexibility in viewer positioning and scalability~\cite{wann1995natural}.

In contrast, alternative approaches aim to expand virtual space by leveraging geometric viewpoint correction and perceptual depth cues, rather than relying on specialized light-field hardware~\cite{cruz2023surround, CAVEReview2014, CrossrealitySOTA2023, RDR2023, AAR2020}. CAVE systems provide immersive multi-surface projection but suffer from cost and scalability constraints~\cite{Vasconcelos2019}, while projection-based XR enables single-wall experiences~\cite{Room2Room, RoomAlive, RealityCheck, See2020} but typically requires precise user tracking and supports only individual use~\cite{RDR2023}. Similarly, 3D anamorphic displays use forced perspective to create depth from a fixed “sweet spot,” offering a glasses-free alternative but still constraining the viewer’s position~\cite{linwei2024workflow, DiPaola2015, matsumoto1997}.
These limitations raise the question of whether shared depth perception can be achieved without individualized rendering. In this study, we explore how to evolve the function of widely installed wall-sized displays by leveraging human cognitive compensation, aiming to support collective 3D experiences from a single fixed viewpoint.

\subsection{Cognitive Threshold Estimation}
Numerous studies have investigated how factors such as screen size and environmental richness affect distance perception in extended spaces viewed through large wall displays and VR HMDs~\cite{gaines2020methods, vienne2020depth, plumert2005distance}. Messing et al.~\cite{messing2005distance} examined depth perception in virtual environments, finding that depth underestimation in VR follows a nearly constant factor. Klein et al.~\cite{klein2009measurement} compared different protocols for measuring distance perception in front of a large-screen immersive display, demonstrating that both the timed imagined walking method and verbal estimation method are effective in wall-display settings. On one hand, the abundance of visual detail and contextual cues on a wall-sized screen might enhance the brain’s ability to compensate for viewpoint differences – essentially providing more “fodder” for the normalization of depth perception~\cite{allison2015perceptual}.

The cognitive threshold is widely used to understand how individuals perceive specific stimuli in virtual environments~\cite{kim2022configuration}. One common method is the two-alternative forced choice (2-AFC) task, where participants choose between two options to indicate their perception of a stimulus, helping determine sensitivity thresholds~\cite{fechner1948elements, jogan2014new}. To overcome the repetitive nature of 2-AFC tasks while maintaining accurate threshold estimates, methods such as the staircase method and the maximum likelihood procedure (MLP) are often employed. The staircase method adjusts stimulus intensity based on participants’ responses to determine the point at which the stimulus is consistently detected~\cite{cornsweet1962staircase, dixon1948method}. MLP optimizes data collection by estimating sensory thresholds based on observed responses~\cite{green1966signal, harvey1986efficient}. In this study, we employ the verbal response method to estimate distance perception for wall displays and use the 2-AFC questionnaire in combination with the staircase method to measure angular thresholds.

\section{Methodology}

\subsection{Research Questions and Hypotheses}

This study aims to leverage human depth perception to allow users to perceive a single image displayed on a large wall as an extension of their physical surroundings. We define this illusion as an \textit{extended virtual space}, and refer to the overall perceptual environment—including both the image and its physical setup—as a \textit{wall display-based XR space}. We investigate how two spatial variables affect user perception: virtual depth ($d_{v}$), the distance between the frontmost part of the virtual object in the XR space and the physical wall display; and absolute viewing distance ($d_{a}$), the distance between the user and the frontmost part of the virtual object in the XR world. It is calculated as the sum of $d_v$ and the physical distance between the user and the wall display ($d_p$), (i.e., $d_a = d_p + d_v$). Our focus is on how changes in these variables influence depth perception and the overall viewing experience in scenarios where users observe artworks in this XR space. Our goal is to identify perceptual thresholds and design principles for enabling multiple users to simultaneously experience depth perception without head-tracking or individualized rendering. To this end, we employ a fixed-perspective image rendered from a universal viewpoint (eye height = $1.5\,\mathrm{m}$), enhanced with vanishing points, shadows, and lighting cues.

We evaluate the system using three perceptual metrics. First, we measure \textit{perceived distance difference (PDD)}, defined as the discrepancy between the system’s intended virtual depth (camera-to-virtual wall distance in the universal setting) and the distance estimated by users. This metric reflects how accurately users perceive depth relative to the intended $d_{a}$. Second, we assess the \textit{viewing angle threshold (VAT)}, the horizontal angle beyond which users begin to lose the sense of depth perception. Participants first viewed the image from the center (i.e., the intended sweet spot), then gradually changed their angle until a noticeable decline in either perceptual factor was reported. Third, we evaluate perceived presence, measured through three subcomponents: realness (authenticity of the scene), spatial presence (sense of “being there”), and visibility (clarity and peripheral awareness). These metrics capture how presence is affected by proximity to the wall, where screen resolution, brightness, and limited peripheral vision may reduce presence. 
~\\
From these constructs, we address the following research questions:

\begin{enumerate} [label={RQ\arabic*}.,noitemsep] 
    \item How does PDD vary as virtual depth ($d_{v}$) and absolute viewing distance ($d_{a}$) change? 
    \item How does VAT vary as a function of $d_{v}$ and $d_{a}$? 
    \item How is perceived presence affected by changes in $d_{v}$ and $d_{a}$? 
\end{enumerate}

For the first hypothesis (H1), prior studies suggest that users tend to underestimate depth as viewing distances increase~\cite{messing2005distance, klein2009measurement, sweet2011depth}. We therefore expect PDD to increase as either $d_{v}$ or $d_{a}$ increases. The second hypothesis (H2) is based on how binocular disparity contributes to depth perception~\cite{howard2002depth, gregory2015eye}. As $d_{v}$ increases, the visual mismatch between the wall-projected image and the viewer’s depth cues becomes more pronounced, potentially narrowing the range in which depth perception is maintained. Thus, we hypothesize that VAT will decrease with increasing $d_{v}$. The third hypothesis (H3) is grounded in presence literature~\cite{PresenceSurvey_CHI_2019, PresenceSurvey_CHI_2024, PQ1998, IPQ2001, slater1994depth}. When the $d_{v}/d_{a}$ ratio increases (i.e., as users move closer to the display), the system becomes more likely to appear artificial due to visible screen resolution, brightness, and reduced peripheral vision~\cite{Lombard1997, Wiederhold2014, Vater2022}, which can diminish perceived presence. 

We therefore propose the following hypotheses:

\begin{enumerate} [label={H\arabic*.},noitemsep] 
    \item PDD will increase with: H1-1) increasing $d_{v}$ and H1-2) increasing $d_{a}$. 
    \item VAT will decrease as $d_{v}$ increases. 
    \item As the ratio $d_{v}/d_{a}$ increases, H3-1) realness ($p_{\text{real}}$), H3-2) spatial presence ($p_{\text{sp}}$), and H3-3) visibility ($p_{\text{vis}}$) will decline, leading to reduced perceived presence within wall display-based XR space. \end{enumerate}

\subsection{Experimental Setup and Apparatus}

\begin{figure}[ht]
 \captionsetup{belowskip=-10pt}
 \centering
 \includegraphics[width=\columnwidth]{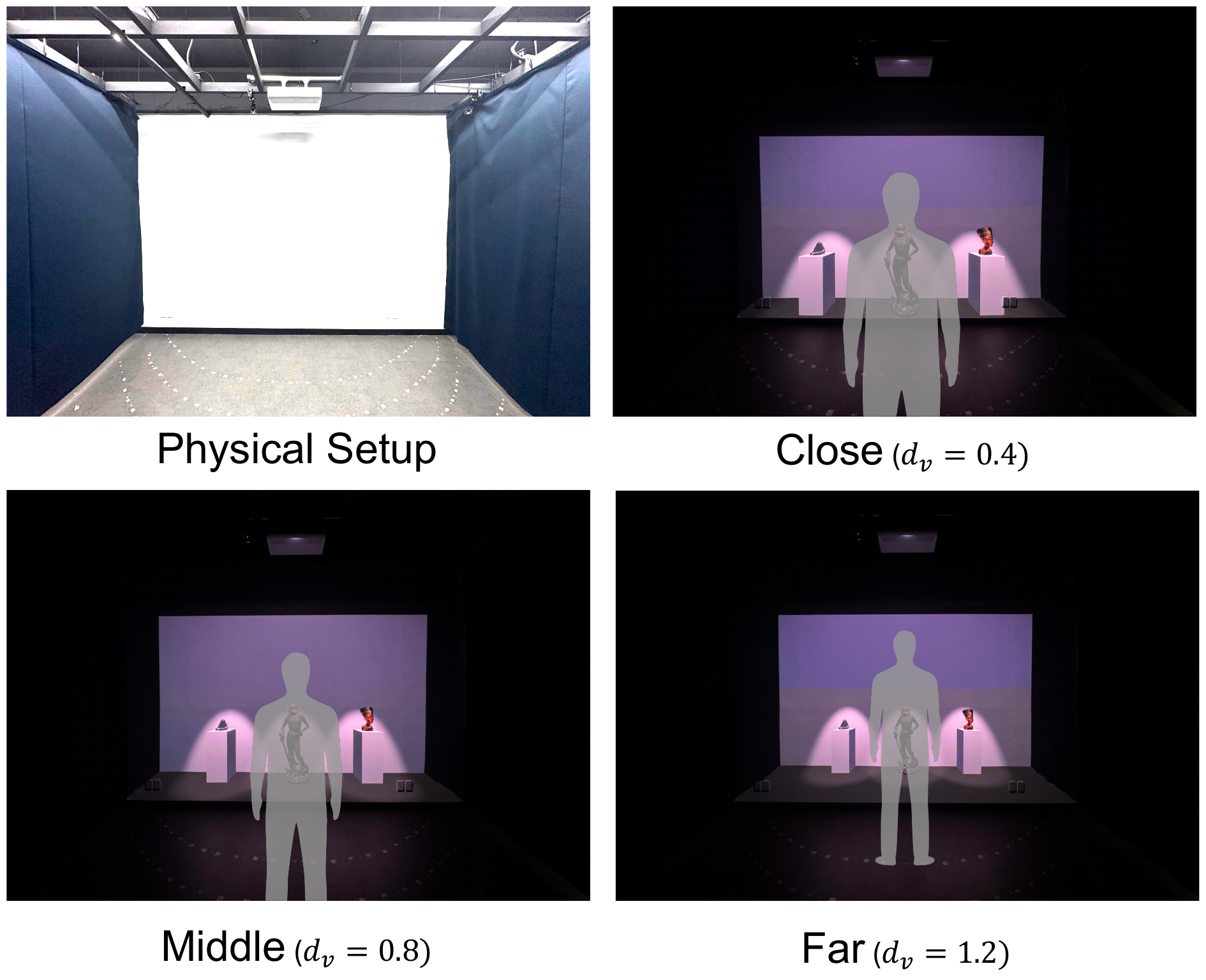}
 \caption{The physical setup of a wall display-based XR space and the three $d_{v}$ conditions (\textit{Close}, \textit{Middle}, \textit{Far}) when $d_{a} = 2\,m$.}
 \label{fig:setup}
\end{figure}

\begin{figure*}[ht]
 \captionsetup{belowskip=-10pt}
 \centering
 \includegraphics[width=\linewidth]{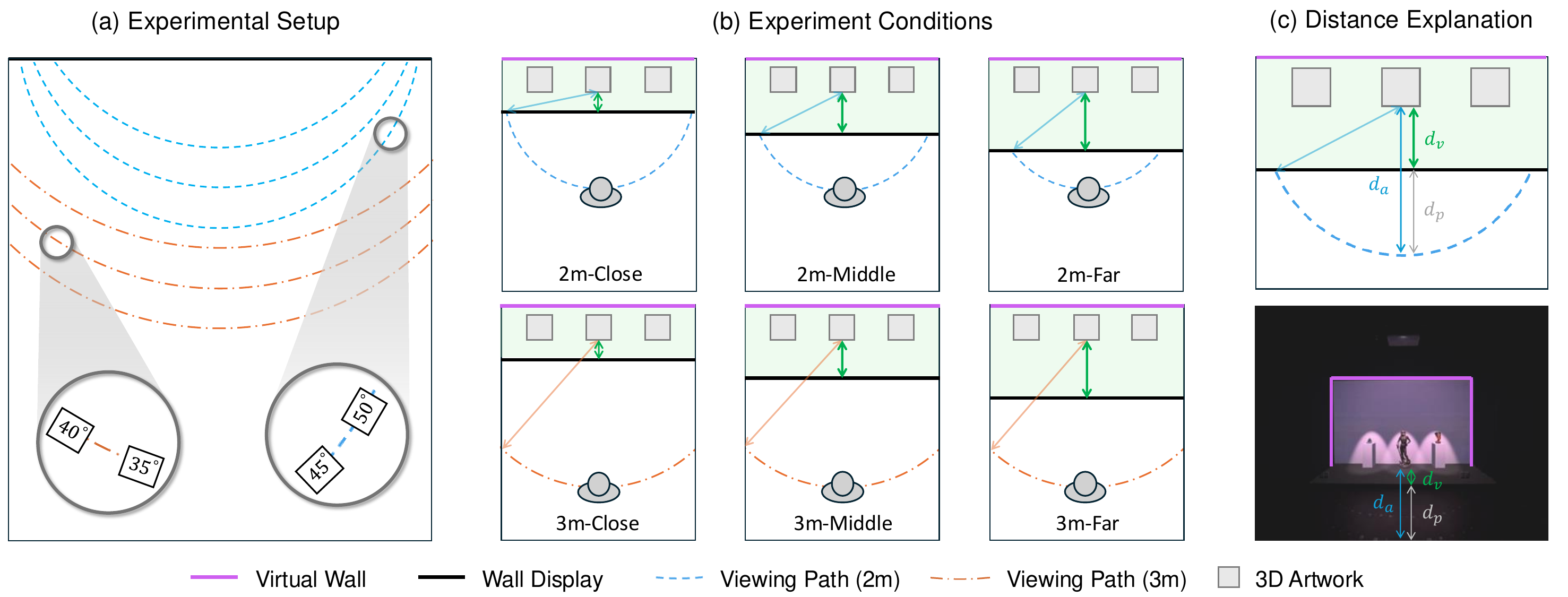}
 \caption{(a) Top view of an experimental setup with six viewing paths with corresponding angles drawn on the floor for VAT estimation, (b) six experimental conditions, and (c) distance explanation ($d_{a}$ = $d_{v}$ + $d_{p}$).}
 \label{fig:conditions}
\end{figure*}

As illustrated in \cref{fig:setup}, we established an experimental environment featuring a large wall display using a single short-throw projector (Samsung The Premiere 9). The projector was ceiling-mounted to illuminate the lower portion of the projection wall fully, and blackout curtains were installed along the edges of the screen to create the illusion that the display extended beyond its physical boundaries, enhancing the immersive effect. The projector supported a 4K resolution (3840$\times$2160) with a throw ratio of 0.189 and a maximum image width of $3.3\,\mathrm{m}$. In our setup, the projected image spanned approximately $4.2\,\mathrm{m}$ in width and $2.36\,\mathrm{m}$ in height. Due to this scaling, the effective resolution was approximately 78.6\% of full 4K. The display setup was monoscopic; therefore, depth perception cues from binocular disparity were not present. The floor area measured $4.2\,\mathrm{m}$ by $4\,\mathrm{m}$ and was enclosed by blackout curtains to minimize visual distractions, as shown in \cref{fig:setup}. All ambient lighting was turned off, and illumination relied solely on the projector output to maintain visual consistency. 

 The extended virtual space content was implemented in Unity (version 2022.3.33f1) and rendered using an Apple M3 Max MacBook Pro 16. To support the perception of an extended virtual space by multiple users, the content was rendered from a fixed viewpoint centered horizontally on the screen. For each condition, this virtual camera was positioned to correspond with the universal viewpoint for a given trial, creating the static, universally rendered image that participants observed from their respective physical locations. The scene was illuminated by a single, default directional light to maintain consistent and neutral shading across all conditions. The height of a virtual camera was positioned at a universal eye height of $1.5\,\mathrm{m}$ to represent a shared viewing perspective. This setup ensured that, when viewed from the central position, the virtual space visually aligned with the physical environment, forming a coherent wall display-based XR space.

To estimate the VAT, six concentric arcs were drawn on the floor to represent different viewing paths. Each arc was centered at the front center point of the stand on which the central virtual object was placed. As illustrated in \cref{fig:conditions}(b), these arcs correspond to the six spatial conditions tested in the experiment. The arcs were grouped into two sets based on absolute viewing distance $d_a$ (\textit{2\,m} and \textit{3\,m}), with each set incorporating three virtual depth ($d_v$) levels—\textit{Close}, \textit{Middle}, and \textit{Far}. Because the virtual wall position changed with $d_v$, the center point of each arc also shifted accordingly. This design ensured that participants experienced consistent angular sampling across all conditions while maintaining alignment with the intended viewing perspective.

\begin{table*}
  \caption{Mean and standard deviation of \textit{perceived distance difference} (PDD), \textit{viewing angle threshold} (VAT), and presence factors.}
  \label{tab:ALL}
  \scriptsize
  \centering
  \begin{tabu} to \linewidth {l c c | c | c c | c c c}
    \toprule
    Condition & $d_{a}$ & $d_{v}$ & PDD [m] & VAT-Left [°] & VAT-Right [°] & $p_{\text{real}}$ & $p_{\text{sp}}$ & $p_{\text{vis}}$ \\
    \midrule
    \textit{2\,m}-\textit{Close} & 2\,m & 0.4\,m & $-0.31$ (0.48) & 37.30 (8.27) & 37.83 (8.58) & 4.45 (1.25) & 4.12 (1.22) & 5.03 (0.96) \\
    \textit{2\,m}-\textit{Middle} & 2\,m & 0.8\,m & $-0.61$ (0.67) & 36.78 (8.85) & 33.62 (9.89) & 4.46 (1.22) & 4.11 (1.24) & 5.09 (0.97) \\
    \textit{2\,m}-\textit{Far} & 2\,m & 1.2\,m & $-0.88$ (0.89) & 24.34 (10.88) & 23.22 (11.26) & 3.46 (1.76) & 3.32 (1.61) & 4.21 (1.46) \\
    \midrule
    \textit{3\,m}-\textit{Close} & 3\,m & 0.4\,m & $-0.36$ (0.42) & 34.93 (6.95) & 32.37 (7.93) & 5.20 (0.67) & 4.84 (0.91) & 5.06 (0.92) \\
    \textit{3\,m}-\textit{Middle} & 3\,m & 0.8\,m & $-0.54$ (0.60) & 31.05 (8.83) & 31.18 (8.59) & 4.82 (1.00) & 4.76 (0.96) & 5.14 (0.88) \\
    \textit{3\,m}-\textit{Far} & 3\,m & 1.2\,m & $-0.73$ (0.82) & 29.01 (8.57) & 27.11 (10.58) & 4.64 (0.96) & 4.26 (1.11) & 4.96 (0.86) \\
    \bottomrule
  \end{tabu}
\end{table*}

\subsection{Study Design}

We selected two $d_{a}$ of \textit{2\,m} and \textit{3\,m} to represent typical viewing scenarios in real-world environments with wall-sized displays, such as museums and public galleries. The \textit{2\,m} distance represents an `engaged viewing' scenario, where a user steps closer to inspect details of the content. In contrast, the \textit{3\,m} distance represents a `casual or group viewing' scenario, where viewers stand further back to perceive the display as part of the larger space. These distances were chosen based on observations of public installations and were determined in a pilot study to be perceptually distinct while both falling within a range typical for such experiences.
The three levels of $d_{v}$ were chosen to investigate a range of extended space experiences, from subtle to pronounced. The \textit{Close} condition ($d_{v}$ = $0.4\,\mathrm{m}$) represented a shallow virtual extension, intended to establish a baseline for perceptible depth. The \textit{Middle} condition ($d_{v}$ = 0.8\,m) was designed to provide a compelling yet comfortable 3D experience, informed by our pilot observations as a potential perceptual sweet spot. Finally, the \textit{Far} condition ($d_{v}$ = $1.2\,\mathrm{m}$) featured a significant depth extension designed to push the perceptual limits of the shared space experience, allowing us to explore potential degradation in perceived quality and presence.
 

The participant's distance from the wall differs according to the $d_v$ as shown in \cref{fig:setup}. As shown in \cref{fig:conditions}(b), combining the two $d_{a}$ and three $d_{v}$ values resulted in a total of six conditions. In the extended virtual space, three content objects were placed: \textit{Fountain} (Marcel Duchamp), \textit{David} (Donatello), and \textit{Nefertiti Bust} (ancient Egyptian artifact). To ensure that users' gaze was not affected by the height of the content objects, the heights of the objects were adjusted using a 1-meter pedestal, in alignment with \textit{David}'s height, so that users' gaze remained at a consistent level. We followed the display guideline of the exhibition design policy from \cite{Barbieri2018, XD2023}.

\vspace{-0.7\baselineskip}
~\\
\textbf{Perceived Distance Difference (PDD)}
The first metric we aimed to measure was the perceived distance to the object within the extended virtual space to assess how much it deviates from the system-induced distance, $d_{a}$. Participants stood at the 0° position, where they could perceive the most accurate depth perception in each condition, and viewed the extended virtual space and the artworks. Afterward, participants verbally responded by saying how far they perceived the distance to the front of the central object~\cite{klein2009measurement}. To measure the PDD, we informed participants that the distance between the two paths—\textit{2\,m}-\textit{Far} and \textit{2\,m}-\textit{Middle}—was $0.4\,\mathrm{m}$, and based on this reference, asked them how far away the object seemed. Participants verbally responded with their perceived distance to the object within the extended virtual space, and we recorded these verbal estimations to measure the perceived distance. The perceived distance was then used to calculate the PDD by subtracting the $d_{a}$ from the perceived distance.

\vspace{-0.7\baselineskip}
~\\
\textbf{Viewing Angle Threshold (VAT)}
The goal of the VAT estimation is to determine how well the human cognitive ability for depth perception correction operates as the viewing angle changes. Through a pilot study, we confirmed that people could experience a relatively high level of depth perception under 0° of each condition. In the main study, the objective was to find the angular threshold at which participants experience depth perception, using the 0° condition as the baseline where participants perceive the most realistic depth. To achieve this, as shown in \cref{fig:conditions}(b), we marked a semi-circle on the floor based on the front of the central object in the participants’ view, using $d_{a}$ as the radius. As $d_{v}$ increased, the circle's center moved deeper into the display, bringing the participants closer to the wall display. To isolate the effects of static pictorial depth cues, motion parallax was intentionally excluded from the perceptual judgment task. Participants were instructed to remain stationary at each designated viewing position while making their assessment. They only moved between trials when instructed to change their viewing location.

We used the pseudo-2AFC method with staircase procedure~\cite{Zenner:2021:UnityStaircase} to estimate VAT. In each angular condition, participants were asked, “Do you feel the depth perception from the extended virtual space and content through the wall display?” Participants were instructed to base their judgment on the sense of stereoscopy they perceived when standing at the 0° position. Participants were instructed to respond “yes” if they felt the same level of depth perception as when viewing from the 0° position, and “no” if they felt that the depth perception of the extended virtual space and the objects within it had diminished in any way. The maximum angle for threshold measurement was limited to 50° when the physical depth was 2\,m and 40° when it was 3\,m, considering our experiment setup and display size. Angles were sampled in 5° increments, with subsequent angles determined using the staircase method based on participant responses. If participants consistently perceived depth at the maximum angle of a condition for three consecutive trials, the staircase method was terminated, and the threshold was recorded as the maximum value (\textit{e.g.}, 50° for $d_{a} = 2\,\mathrm{m}$ and 40° for $d_{a} = 3\,\mathrm{m}$).

\vspace{-0.7\baselineskip}
~\\
\textbf{Perceived Presence Factors}
The last metric we measured was perceived presence factors—realness, spatial presence, and visibility—to examine how users’ presence differs as $d_{v}$ and $d_{a}$ change in a wall display-based XR space.
Two questionnaires commonly used for evaluating presence in VR environments—the Igroup Presence Questionnaire (IPQ) and the Witmer and Singer Presence Questionnaire (WSPQ)—were utilized \cite{PQ1998, IPQ2001}.
We adapted and tailored items from the IPQ and WSPQ questionnaires to match the perceived presence factors we aimed to measure. Below, we explain the selected factors, the rationale for choosing them, and how the questionnaire items were selected and adapted:

\begin{itemize}
    \vspace{-0.5\baselineskip}
    \item \textbf{Realness ($p_{\text{real}}$)}
    To measure realness in wall display-based XR, it is crucial to assess how seamlessly and consistently the extended virtual space is integrated with the real-world space. For this, we deployed the items from IPQ-REAL, which measures realness. Due to the nature of wall display-based XR, which is connected to the real world, we excluded item 4, which asks whether the environment feels more realistic than the real world. Instead, we modified an item to assess how naturally the extended virtual space appeared when moving during the angular threshold measurement task (WSPQ-5).
    \vspace{-0.5\baselineskip}
    \item \textbf{Spatial Presence ($p_{\text{sp}}$)}
    As extended virtual space is projected through a display screen in wall display-based XR, to measure spatial presence—\textit{a sense of being in a place}, it is important to assess how much users felt physically present in the space, rather than simply perceiving it as a screen. For this, we selected items from IPQ-SP and IPQ-PRES, excluding item 3, a reverse-scored item of 5, in order to maintain a balanced representation across the factors. We also excluded item 4, as our study did not involve any operation tasks.
    \vspace{-0.5\baselineskip}
    \item \textbf{Visibility ($p_{\text{vis}}$)}
    Visibility plays a significant role in contributing to the presence in VR \cite{PQ1998, hvass2017visual, wirth2007process}. Especially with wall displays, which induce more constraints to delivering immersion and presence compared to HMDs, ensuring visibility is even critical \cite{Bimber2002, Shu2018}. To measure visibility, we adapted items from WSPQ that concern visibility factors: visual capability (9), visual clarity (11, 12), and visual display quality (17).
\end{itemize}

\vspace{-0.5\baselineskip}

All questionnaire items were based on a 7-point Likert scale. For each participant, average values for the four items associated with each factor were calculated to derive scores for $p_{\text{real}}$, $p_{\text{sp}}$, and $p_{\text{vis}}$. These scores were obtained for six conditions, and the full list of questions can be found in the table of the supplementary materials.

\subsection{Participants and Task Procedures}

An institutional review board approved the study's content and procedures. We recruited 20 participants (11 male and nine female) through the local university website, and each was paid \$12 in remuneration. All participants were at least 18 years old and had normal or corrected-to-normal vision. The participants' ages ranged from 19 to 33 years, with a mean age of 25.45 (SD = 3.40). Their heights ranged from $157\,\mathrm{cm}$
 to $183\,\mathrm{cm}$, with a mean height of $168.93\,\mathrm{cm}$ (SD = 8.78). Participants were first given a basic explanation of the experiment before entering our wall display-based XR space. The trial task involved viewing a single bust of Augustus artwork on the stand, which is not used for the main study, under \textit{3\,m}-\textit{Middle} ($d_{v} = 0.8\,\mathrm{m}$, $d_{a} = 3\,\mathrm{m}$)
. This task allowed participants to familiarize themselves with verbally responding to the perceived distance and the 2-AFC questionnaire for VAT estimation. After the trial task, participants proceeded to the main task. 

First, participants stood at the center (0°) of the arc corresponding to their assigned condition and verbally reported the perceived distance to the artwork. Then, they were guided to move to the left or right and answer the 2-AFC questionnaire for VAT estimation. After the participant responded with either “yes” or “no” at the given angle, a short sound confirmed the response was recorded, and the next angle was announced via Google Translate's text-to-speech system\footnote{https://translate.google.com/} based on the calculated next angle from the staircase method. Once participants had completed responses in both \textit{Directions} (\textit{Left}, \textit{Right}) in a given condition, they exited the wall display-based XR space to complete the perceived presence questionnaire. This process was repeated for all six conditions, with the order of conditions randomized and counterbalanced using the Latin square method. After completing all tasks, a debriefing interview covered their overall experience, the criteria for determining the loss of depth perception, the impact of virtual and physical depth, and the most preferred condition. The entire experiment took about 45 minutes per participant, and no participants reported experiencing motion sickness or difficulty completing the tasks.

\section{Results}

\begin{figure*}[ht]
 \centering
 \captionsetup{aboveskip= 0pt, belowskip=0pt}
 \includegraphics[width=\linewidth]{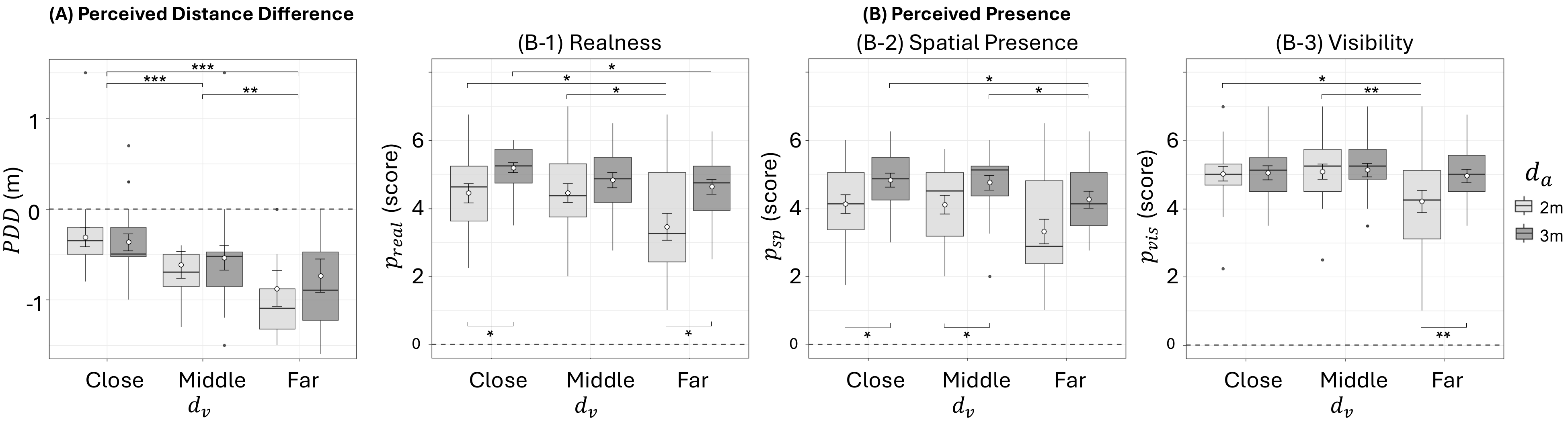}
 \caption{Statistic results of (A) PDD and (B) perceived presence factors (B-1) realness ($p_{\text{real}}$), (B-2) spatial presence ($p_{\text{sp}}$), and (B-3) visibility ($p_{\text{vis}}$). The median (bold lines in the box plot), means (hollow dots), outliers (filled dots) are drawn, error bars representing the standard error of the mean (SEM), and significantly different pairs are marked. ($*: p < .05$; $**: p < .01$; $***: p < .001$)}
 \label{fig:statResults}
\end{figure*}

\subsection{Perceived Distance Difference (PDD)}

\cref{tab:ALL} shows the mean and standard deviation (SD) of the PDDs for each condition, with the corresponding statistical results visualized in \cref{fig:statResults}(A).
Results show that participants tend to underestimate the distance to the object in the extended virtual space as $d_v$ increases. To validate our first hypothesis, the Shapiro-Wilk normality test was conducted for each combination of $d_{a}$ and $d_{v}$ on PDD. Since the assumption of normality is violated across all groups, we used an Aligned Rank Transform (ART)~\cite{wobbrock2011aligned} analysis to evaluate the effects of $d_{a}$ and $d_{v}$ on PDD, considering the repeated measures design. The analysis revealed a significant main effect of $d_{v}$, $F(2, 95) = 26.71$, $p < .001$. However, there was no significant main effect of $d_{a}$, $F(1, 95) = 2.40$, $p = .125$, nor a significant interaction effect between the two factors, $F(2, 95) = 1.79$, $p = .173$.

Post-hoc pairwise comparisons, using Wilcoxon signed-rank tests with a Bonferroni correction, were conducted to examine the significant main effect of $d_v$.  As shown in \cref{fig:statResults}(A), the analysis revealed a significant difference between all three conditions in both $d_a$ conditions. The PDD in the \textit{Middle} condition was significantly different from the \textit{Close} condition ($p < .001$, $r = .66$). Similarly, the \textit{Far} condition was significantly different from the \textit{Close} condition ($p < .001$, $r = .65$). A significant difference was also found between the \textit{Middle} and \textit{Far} conditions ($p = .008$, $r = .50$). These results indicate that increasing the virtual depth at each level led to a significant change in participants' depth perception.
All significant comparisons yielded large effect sizes ($r > .5$) with a mean of .60, indicating consistent and practically meaningful differences between conditions.

\subsection{Viewing Angle Threshold (VAT)}

\captionsetup{aboveskip= 8pt, belowskip=-5pt}
\begin{figure}[h]
 \centering
\includegraphics[width=\columnwidth]{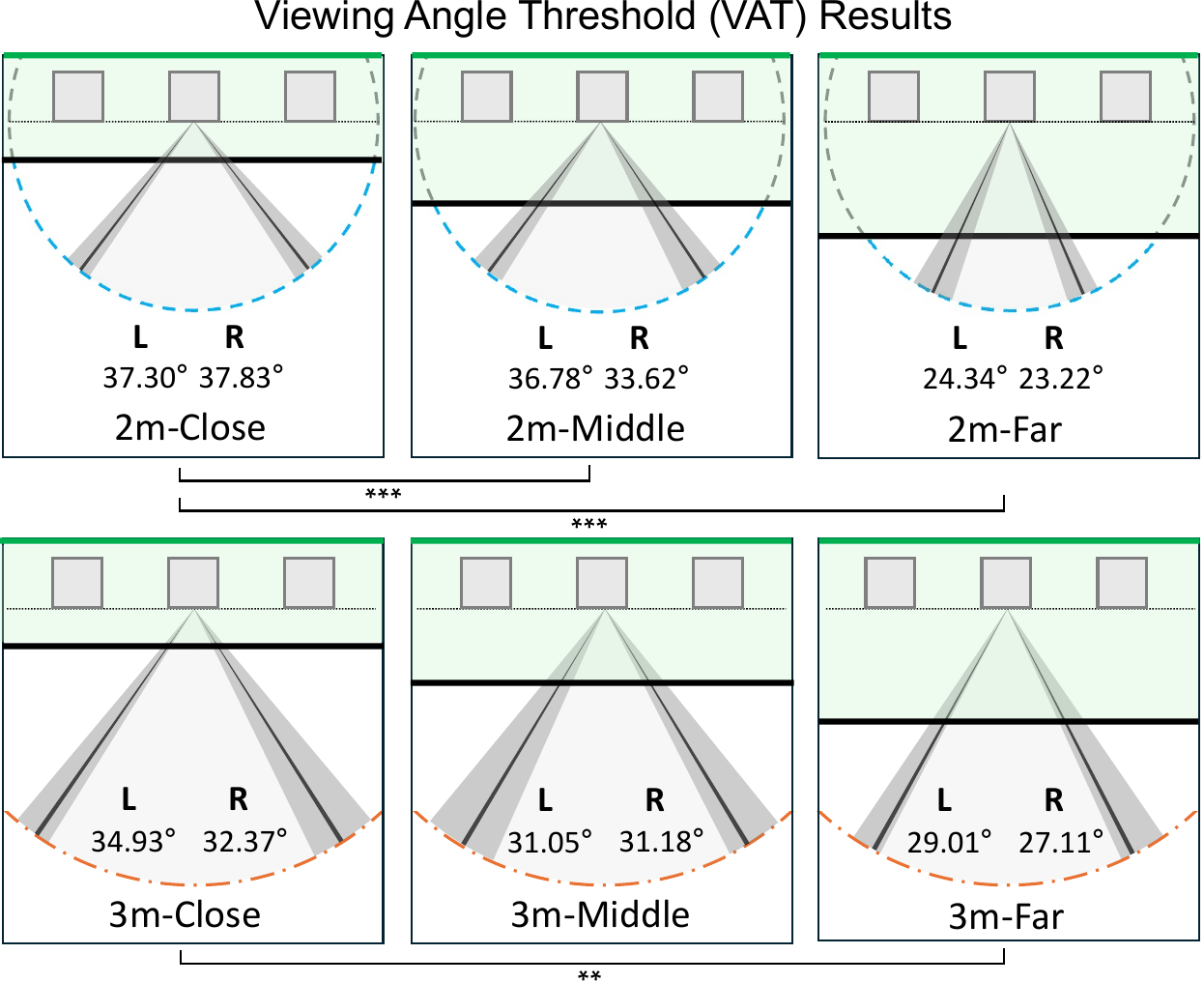}
 \caption{Mean VAT values of left (L) and right (R) from top views of six experimental conditions. ($**: p < .01$; $***: p < .001$)
 }
 \label{fig:vat}
\end{figure}

\begin{table*}[t]
\centering
\caption{Summary of statistical results for Viewing Angle Threshold (VAT) under \textit{2\,m} and \textit{3\,m} $d_a$ conditions.}
\label{tab:vat_summary}
\begin{tabular}{l l l}
\toprule
\textbf{Effect / Comparison} & \textbf{$d_a$ = \textit{2\,m} (Repeated Measures ANOVA)} & \textbf{$d_a$ = \textit{3\,m} (ART ANOVA)} \\
\midrule
\multicolumn{3}{l}{\textit{\textbf{Main and Interaction Effects}}} \\
\quad $d_v$ (Virtual Depth) & $F(2, 36) = 19.63$, $p < .001$, $\eta_g^2 = .29$ *** & $F(2, 90) = 8.80$, $p < .001$ *** \\
\quad Direction & $F(1, 18) = 1.59$, $p = .223$, $\eta_g^2 = .004$ & $F(1, 90) = 1.19$, $p = .279$ \\
\quad $d_v \times$ Direction & $F(2, 36) = 1.44$, $p = .251$, $\eta_g^2 = .006$ & $F(2, 90) = 0.16$, $p = .851$ \\
\midrule
\multicolumn{3}{l}{\textit{\textbf{Post-hoc Pairwise Comparisons for $d_v$}}} \\
\quad Close vs. Middle & $t(37) = -7.17$, $p_{\text{adj}} < .001$, $d = -1.16$ *** & $V = 222$, $p_{\text{adj}} = .25$, $r = .29$ \\
\quad Close vs. Far & $t(37) = -6.56$, $p_{\text{adj}} < .001$, $d = -1.06$ *** & $V = 116$, $p_{\text{adj}} = .001$, $r = .58$ ** \\
\quad Middle vs. Far & $t(37) = -1.37$, $p_{\text{adj}} = .54$, $d = -0.22$ & $V = 91$, $p_{\text{adj}} = .097$, $r = .38$ \\
\bottomrule
\end{tabular}
\end{table*}

\cref{tab:ALL} shows each condition's mean and SD of VAT. Results show that the range of angular thresholds that could support depth perception was distributed from about 23° to 37°. To analyze the effects of our independent variables on the VAT, we conducted separate analyses for each $d_a$. As the normality assumption was met for the \textit{2\,m} data but violated for the \textit{3\,m} data, a two-way repeated measures ANOVA and a non-parametric ART ANOVA were used, respectively. The detailed statistical results of these analyses and the corresponding post-hoc tests are summarized in \cref{tab:vat_summary}.

\cref{fig:vat} illustrates the mean VAT for each condition, with statistically significant pairwise differences marked at the bottom. The analyses revealed a consistent and significant main effect of $d_v$ on the VAT for both the \textit{2\,m} and \textit{3\,m} conditions. Conversely, there was no significant main effect of Direction, nor a significant interaction effect between the two factors in either condition. Post-hoc tests on the significant main effect of $d_v$ provided further insight. For the \textit{2\,m} distance, the \textit{Close} condition produced a significantly different VAT from both the \textit{Middle} and \textit{Far} conditions. For the \textit{3\,m} distance, the difference was most pronounced between the \textit{Close} and \textit{Far} conditions. This suggests that while shallower virtual depths consistently allow for a wider range of viewpoint tolerance, the specific perceptual thresholds can shift with viewing distance.
All significant comparisons yielded large effect sizes ($r > .5$, and $|d| > .8$), indicating practically meaningful differences between conditions.

\subsection{Perceived Presence Factors}

\begin{table*}[t]
\centering
\caption{Post-hoc pairwise comparison results for presence factors ($p_{\textbf{real}}$, $p_{\textbf{sp}}$, $p_{\textbf{vis}}$). Results that meet the adjusted significance threshold for each comparison type ($p_{\text{adj}} < .017$ for $d_v$ comparisons within \textit{2\,m} or \textit{3\,m}, and $p_{\text{adj}} < .025$ for $d_a$ (\textit{2\,m} vs. \textit{3\,m}) comparisons) are in \textbf{bold}.
}
\label{tab:presence_posthoc}
\begin{tabular}{l l c c c}
\toprule
\textbf{Comparison Type} & \textbf{Pairwise Comparison} & \textbf{$p_{\text{real}}$} ($p_{\text{adj}}$, $r$) & \textbf{$p_{\text{sp}}$} ($p_{\text{adj}}$, $r$) & \textbf{$p_{\text{vis}}$} ($p_{\text{adj}}$, $r$) 
\\
\midrule
\multirow{3}{*}{\textbf{Within 2\,m} ($d_v$ comparison)} 
& Close vs. Middle & .729, .06 & .521, .10 & .716, .06 \\
& \textbf{Close vs. Far} & \textbf{.006, .44} & .023, .36 & \textbf{.006, .44} \\
& \textbf{Middle vs. Far} & \textbf{.008, .42} & .036, .33 & \textbf{.002, .48} \\
\midrule
\multirow{3}{*}{\textbf{Within 3\,m} ($d_v$ comparison)} 
& Close vs. Middle & .041, .32 & .703, .06 & .499, .11 \\
& \textbf{Close vs. Far} & \textbf{.007, .43} & \textbf{.009, .41} & .403, .13 \\
& \textbf{Middle vs. Far} & .187, .21 & \textbf{.009, .42} & .286, .17 \\
\midrule
\multirow{3}{*}{\textbf{Between 2\,m vs. 3\,m} ($d_a$ comparison)} 
& at Close & \textbf{.016, .37} & \textbf{.015, .38} & .897, .02 \\
& at Middle & .192, .21 & \textbf{.008, .42} & .868, .03 \\
& at Far & \textbf{.004, .45} & .028, .35 & \textbf{.003, .47} \\
\bottomrule
\end{tabular}
\end{table*}

\cref{tab:ALL} shows the mean and standard deviation of the presence factors for each condition.
Since the data violated the assumption of normality (Shapiro-Wilk test, $p < .05$), we conducted a non-parametric Friedman test for each presence factor. 
The test revealed significant differences for all three factors: \textbf{$p_{\text{real}}$}, $\chi^2$(5) = 59.31, p $<$ .001; \textbf{$p_{\text{sp}}$}, $\chi^2$(5) = 62.86, p $<$ .001; \textbf{$p_{\text{vis}}$}, $\chi^2$(5) = 64.97, p $<$ .001.
To investigate these differences, post-hoc Wilcoxon signed-rank tests with a Bonferroni correction were performed. The detailed results of all pairwise comparisons are summarized in \cref{tab:presence_posthoc}. 

A consistent pattern emerged indicating that the \textit{Far} condition significantly reduced presence factors compared to the \textit{Close} condition. 
This effect was particularly pronounced for $p_{\textbf{real}}$ across both $d_a$, for $p_{\textbf{sp}}$ at the \textit{3\,m} distance, and for $p_{\textbf{vis}}$ at the \textit{2\,m} distance. 
Additionally, the \textit{Far} condition yielded significantly lower presence factor scores than the \textit{Middle} condition, specifically for $p_{\textbf{real}}$ and $p_{\textbf{vis}}$ within the \textit{2\,m} distance and for $p_{\textbf{sp}}$ within the \textit{3\,m} distance. 
Direct comparisons between $d_a$ conditions further revealed that the \textit{2\,m} condition resulted in significantly lower presence factors than the \textit{3\,m} condition at \textit{Close} ($p_{\textbf{real}}$ and $p_{\textbf{sp}}$), \textit{Middle} ($p_{\textbf{sp}}$), and \textit{Far} ($p_{\textbf{real}}$ and $p_{\textbf{vis}}$) conditions.
The effect sizes of statistically significant pairwise comparisons ranged from medium ($r \approx .37$) to large ($r \approx .48$), with a mean effect size of $r = .43$, indicating consistent and practically meaningful differences between conditions.

\subsection{Interview}
\textbf{Perceptual Realness}
First, participants said they felt depth perception through realistic representation in displayed images. They mentioned the following as realistic representations: structural methods through the size of the artwork for representing $d_{v}$, light and shadow through lighting, texture, color contrast through the resolution, and saturation of the artwork on display (P5:\textit{“The shadows cast on the cube that the artwork was placed on looked three-dimensional when it made structural sense.”}), (P1,8:\textit{“The statue in the center has a good resolution and realistic reflections from the light, giving it a three-dimensional feel.”}). 


They also centered on the immersion and atmosphere that virtual spaces projected on the wall create because of the natural connection to the physical space (P5:\textit{“I realized that I was immersed when I saw the connection between the projected screen and the curtain.”}), (P19:\textit{“When We visit a museum, we're surrounded by the artwork, and this similar view gives me a real museum experience.”}). Based on interview responses, we confirmed that users could not only perceive a sense of depth perception, but our design also effectively extended the exhibition space in a way that felt realistic to the users.

\vspace{-0.7\baselineskip}
~\\
\textbf{Extended Space Experience} 
Participants observed that as $d_{v}$ decreased in the \textit{Close} condition, the range of angles from which they could perceive depth extended (P13:\textit{“As the virtual depth decreased, I felt that the range of angles where depth perception was possible extended.”}). Particularly, they specifically noted that their experience suddenly felt different at certain angles when responding to the perception of depth (P8:\textit{“As the angle changed, I strongly felt that the space suddenly became distorted.”}). As $d_{v}$ increased, participants experienced more distortion due to differences in their line of sight, and the sense of depth diminished (P10:\textit{“As the virtual depth increased, when looking downwards from a distance, the sense of depth disappeared.”}), (P12,16,20:\textit{“When the display was positioned lower and not aligned with eye level, the feeling of distortion intensified.”}).


Additionally, participants emphasized that visibility is crucial for accurately understanding the artwork, highlighting the importance of maintaining an appropriate distance, along with depth perception, during the viewing experience (P9:\textit{“They felt that the object was too far away for proper viewing, making the \textit{Middle} condition the most preferred.”}), (P20:\textit{“When I stood too far away, it felt like I couldn’t view the exhibition properly. I preferred the middle distance."}), (P13:\textit{“If there’s no significant difference in depth perception, I preferred Middle because I could see the details of the artwork better.”}), (P3:\textit{“While greater distance increased immersion, it made it harder to explore the artwork.”}). As $d_{v}$ decreased, participants indicated enhanced depth perception and immersion; however, they found it more difficult to examine the artwork closely. For this reason, most participants favored the \textit{Middle} condition.

\vspace{-0.7\baselineskip}
~\\
\textbf{Immersion in Virtual Space} 
Within the space we designed, participants could experience an extended sense of space as $d_{v}$ decreased, blending the boundaries between the real and virtual environments (P10:\textit{“As the physical depth increased, the sense of spatial extension became more apparent, and the objects felt more three-dimensional.”}), (P2:\textit{“While angle is important, the greater viewing distance, the stronger the immersion. It felt similar to being affected by distance in an immersive screen or movie.”}), (P16:\textit{“The further the physical depth, the more I felt like I was standing in a virtual space.”}). Conversely, when $d_{v}$ increased, participants reported a reduction in immersion due to the awareness of the screen and surrounding walls (P7:\textit{“When the physical depth was closer, I felt a greater disconnection between the virtual wall distance, the person standing, and the edges of the projected wall.”}), (P18:\textit{“The further the virtual depth, the narrower my field of view became, making it harder to feel a sense of space and reducing immersion."}). 

Additionally, in the \textit{2\,m}-\textit{Far} condition, participants felt that the wall display was too close, leading to an easily perceive the extended space as flat (P13:\textit{“When I am close, the flatness of the display becomes more noticeable, and I can feel the immersion break significantly.”}), (P14,17:\textit{“When the physical depth was close, it felt more like looking at a screen, but when it was farther, it felt like I was in an actual museum.”}). Through this, we confirmed that our extended space configuration, connected with the projected 2D image, provided an extended sense of space and immersion.

\section{Discussion}

\subsection{Analysis}

We accept H1-1, which hypothesized that as the $d_{v}$ increases, participants would perceive objects in the extended space through a large wall display as being closer than the system intended. The results support this hypothesis, as all three combinations of $d_{v}$ had a significant effect on PDD. Furthermore, the post-hoc analysis of the interaction effect between $d_{v}$ and $d_{a}$ showed that all three $d_{v}$ pairs within each $d_{a}$ condition had a significant effect. There are two main interpretations for why users tend to underestimate the distance to objects in the space as virtual depth increases. First, human depth perception tends to be more accurate at closer distances, but the estimation error increases as objects become farther away~\cite{plumert2005distance}. The human brain processes depth by interpreting the slight differences in the images perceived by each eye. However, when viewing a wall display, both eyes receive the same image, creating discrepancies. Additionally, observing the space from an angle that differs from the user's eye level introduces depth distortion, which likely explains why the error increases as depth grows. Our results confirm that when participants view relatively close objects through a large wall display, $d_{v}$ significantly impacts PDD.


We reject H1-2, which hypothesized that as the $d_{a}$ increases, participants would perceive objects in the extended space through a large wall display as being closer than the system intended. The two $d_{a}$ conditions did not have a significant effect on PDD, and even in the additional post-hoc analysis for the interaction effect between $d_{a}$ and $d_{v}$. As shown in \cref{tab:ALL}, participants tended to underestimate depth in all conditions. This aligned with Klein et al.~\cite{klein2009measurement}'s results that the perceived distance tends to decrease as the $d_{a}$. However, since the two $d_{a}$ conditions (\textit{2\,m} and \textit{3\,m}) in our study were relatively close compared to Klein's study investigated from 2\,m to 15\,m, no significant difference between these two conditions was observed. Based on the results, we conclude that when $d_{a}$ is relatively close, $d_{a}$ is not a significant factor for judging the distance to the object through the wall display.

We conditionally accept H2, which hypothesized that as $d_{v}$ increases, the VAT at which participants perceive depth from the extended virtual space would decrease. As shown in \cref{tab:ALL}, VAT tends to decrease as $d_{v}$ increases. However, statistically significant differences were observed only between \textit{2\,m}-\textit{Close} vs. \textit{2\,m}-\textit{Middle}, \textit{2\,m}-\textit{Close} vs. \textit{2\,m}-\textit{Far}, and \textit{3\,m}-\textit{Close} vs. \textit{3\,m}-\textit{Far}. The primary reason for this result appears to be that as $d_{v}$ increases, the difference between the images seen by the two eyes on a 2D display and the images they see when viewing real objects in a physical space with depth becomes larger. 
The reason for the significant difference only observed between \textit{Close} and \textit{Middle} in $d_{a}$ = \textit{2\,m} condition is attributed to the fact that when $d_{a}$ = \textit{2\,m}, the $d_{p}$ is shorter than the $d_{a}$ = 3\,m condition by 1\,m, results in more pronounced image distortion because participants get closer to the display as the angle increases. On the other hand, the non-significant results were found between \textit{Close} and \textit{Middle} across both $d_{a}$ conditions. It suggests that the effect of $d_{v}$ on VAT is relatively small within the $d_{v}$ range from 0.4\,m to 0.8\,m in our experimental setup. 

We conditionally accept H3-1, which hypothesizes that as the ratio of $d_{v}$ to $d_{a}$ increases (i.e., as user getting close to the wall display) in wall display-based XR space, realness worsens. When $d_{a}$ was $2\,\mathrm{m}$, the \textit{Far} condition showed significantly lower $p_{\text{real}}$ values compared to both \textit{Middle} and \textit{Close}, indicating that realness decreases as $d_{v}$ increases. 
Although the \textit{Middle} condition had a slightly higher mean value than the \textit{Close} condition, no statistically significant difference was observed. As supported by interview results, participants showed a preference for the \textit{Middle} $d_{v}$, which may interpret the case.
When $d_{a}$ was \textit{3\,m}, the \textit{Far} condition again had significantly lower $p_{\text{real}}$ compared to \textit{Close}, reflecting a similar pattern as at \textit{2\,m}. However, no significant differences were found between \textit{Middle} and \textit{Far}, or \textit{Close} and \textit{Middle}, possibly due to the smaller interval of ratio of $d_{v}$ to $d_{a}$ as the distance increased from \textit{2\,m} to \textit{3\,m}.
Based on comparisons between \textit{Middle} and \textit{Close} in both $d_{a}$, we infer that when the ratio of $d_{v}$ to $d_{a}$ is less than or around half, $d_{v}$ does not significantly affect realness.
In the comparisons of $d_{v}$ across the \textit{2\,m} and \textit{3\,m} conditions, significant differences were observed between \textit{2\,m}-\textit{Close} vs. \textit{3\,m}-\textit{Close}, as well as \textit{2\,m}-\textit{Far} vs. \textit{3\,m}-\textit{Far}, with $p_{\text{real}}$ values being significantly lower in the \textit{2\,m} conditions. These results further support hypothesis H3-1, indicating that realness decreases as the ratio of $d_{v}$ to $d_{a}$ increases.

We conditionally accept H3-2, which hypothesizes that as the ratio of $d_{v}$ to $d_{a}$ increases in wall display-based XR space, spatial presence worsens. When $d_{a}$ was \textit{3\,m}, the \textit{Far} condition showed significantly lower $p_{\text{sp}}$ values compared to both \textit{Middle} and \textit{Close}, indicating that spatial presence decreases as $d_{v}$ increases.
In comparisons of $d_{v}$ across the \textit{2\,m} and \textit{3\,m} conditions, significant differences were observed between \textit{2\,m}-\textit{Close} vs. \textit{3\,m}-\textit{Close}, as well as \textit{2\,m}-\textit{Middle} vs. \textit{3\,m}-\textit{Middle}, with $p_{\text{sp}}$ values being significantly lower in the \textit{2\,m} conditions than in the \textit{3\,m} conditions. These results further support the hypothesis H3-2, indicating that spatial presence decreases as the ratio of $d_{v}$ to $d_{a}$ increases.
In contrast, when comparing $d_{v}$ conditions within the \textit{2\,m}, no statistically significant differences were found. Interview results—\textit{“When close to the wall display, it felt more like looking at a screen rather than feeling a sense of being in a place.”}—suggest that both the $d_{v}$ conditions in \textit{2\,m} and the \textit{3\,m}-\textit{Far} condition were perceived as too close to the screen, leading to a lack of spatial presence.

We conditionally accept H3-3, which hypothesizes that as the ratio of $d_{v}$ to $d_{a}$ increases in wall display-based XR space, visibility worsens. When $d_{a}$ was $2\,m$, the \textit{Far} condition showed significantly lower $p_{\text{vis}}$ values compared to both \textit{Middle} and \textit{Close}, indicating that visibility decreases as $d_{v}$ increases.
In comparisons of $d_{v}$ across the \textit{2\,m} and \textit{3\,m} conditions, significant differences were observed between \textit{2\,m}-\textit{Far} vs. \textit{3\,m}-\textit{Far}, with $p_{\text{vis}}$ values being lower in the \textit{2\,m} conditions than in the \textit{3\,m} conditions. These results further support hypothesis H3-3, indicating that visibility decreases as the ratio of $d_{v}$ to $d_{a}$ increases.
In contrast, no statistically significant differences were found when comparing $d_{v}$ conditions within the \textit{3\,m}, or between the \textit{Middle} and \textit{Close} conditions across both \textit{2\,m} and \textit{3\,m}.
This can be attributed to the fact that, as indicated in the interview results, in extreme cases like \textit{2\,m}-\textit{Far} where $d_{v}$ exceeds half of $d_{a}$, the field of view (FoV) narrows, leading to reduced visibility.
Taking all these findings into account, we conditionally accept H3, which hypothesizes that as the ratio of $d_{v}$ to $d_{a}$ increases, perceived presence within wall display-based XR space decreases.

\subsection{Design Implications for Wall Display-based XR Space}

This study explored the potential of a wall display as a medium for virtual space extensions by leveraging human cognitive compensation to present an extended virtual space for multi-user co-viewing. Based on our findings in the context of exhibition-oriented space extension, we propose the following design implications for constructing wall display-based XR space.

\textbf{Calibrating content placement to account for depth underestimation.} The goal of wall display-based XR space is to enable multiple users to experience a consistent level of depth perception without requiring wearable devices. Because both eyes view the same image on the screen, participants tended to underestimate perceived depth more than in prior studies using stereoscopic glasses. As shown in \cref{tab:ALL}, objects were consistently perceived as being closer than their intended positions, with this underestimation becoming more pronounced as $d_{v}$ increased. Therefore, when placing content within an extended virtual space, designers may consider positioning virtual objects slightly farther back than mathematically intended, particularly in static scenarios such as 3D exhibitions. This calibration may better convey the intended sense of depth.

\textbf{Using viewing angle thresholds to define depth perception zones.} A key contribution of our study is the empirical identification of what we term a `zone of viewpoint tolerance' — found to be between 23° and 37° under our specific experimental conditions. This finding should not be interpreted as a universal perceptual law, but rather as the first quantitative baseline for a shared, non-tracked XR experience. This provides a crucial, practical guideline for designers of public installations. For a given rendered viewpoint on a wall-sized display, our results suggest that a compelling 3D experience can be successfully maintained for multiple, off-axis users without requiring individual head-tracking, as long as they are positioned within this significant viewing cone.

One notable finding was that participants experienced a convincing sense of depth even at larger horizontal viewing angles than expected, despite the image not aligning precisely with their eye level. In interviews, several participants expressed surprise at how immersive the scene felt without head-mounted displays or adaptive rendering. This suggests that wall display-based XR can offer a compelling 3D viewing experience from a range of angles. Furthermore, when asked about the point at which the image “lost its depth perception,” participants often described a sudden drop in depth, rather than a gradual fade. This implies that viewers can integrate visual cues effectively up to a certain threshold, after which depth collapses. As illustrated in \cref{fig:vat}, defining and indicating depth perception zones using VAT data and qualitative feedback may help ensure shared perceptual experiences among multiple users.

\textbf{Balancing the ratio of $d_{v}$ to $d_{a}$ to support presence.} The wall display-based XR space enables users to perceive digital content as if it is seamlessly extended from the physical environment. However, relying solely on visual cues from a single 2D surface means that excessive virtual depth can lead to perceptual distortions, especially for distant objects. While our results showed that realness, spatial presence, and visibility generally worsen as $d_{v}$ increased, interviews revealed a nuanced perspective. Some participants reported that overly deep scenes felt less immersive or more artificial, while others preferred the \textit{Middle} condition, which maintained visual continuity with the physical environment and allowed for better detail recognition. Considering both the depth perception area and overall presence under our conditions, the 3m-middle condition (with $d_{v} = 0.8\,m$ and $d_{a} = 3\,m$) was determined to be the most appropriate. The optimal $d_{v}$ to $d_{a}$ ratio may depend on factors such as screen size, content type, and ambient lighting, but adjusting it in reference to VAT data and content characteristics can enhance overall user experience. Based on this, we suggest that $d_{v}$ be carefully balanced—not too shallow, which can not extend the space enough, and not too deep, which may reduce presence. 

\subsection{Limitations}

While our study presents a preliminary exploration of human depth perception to extended virtual spaces projected on a wall-sized display, several limitations should be acknowledged to guide future research. First, although the system was designed to support multi-user viewing through universal rendering without individual calibration, we did not directly evaluate scenarios involving simultaneous co-viewing by multiple users. In real-world settings, social interactions, shared attention, and different body postures can alter perceptual focus and cognitive load, potentially influencing VAT. Future work could examine how dynamic group interactions affect depth perception in shared XR environments.

Second, while we implemented a holistic set of key visual cues—such as vanishing points, shadows, and gradients—to support seamless perceptual integration, the effects of each cue were not isolated or systematically analyzed. This means the 23°–37° `zone of viewpoint tolerance' we identified should be considered a specific outcome resulting from the particular combination of cues in our stimulus. It is highly probable that this perceptual zone would expand or contract depending on the strength, weakness, or absence of certain cues. Therefore, future research is needed to deconstruct these elements and examine how individual visual factors, such as the quality of shadows or the presence of motion parallax, quantitatively influence the boundaries of viewpoint-tolerant depth perception on wall displays.

Lastly, our system adopts a universal viewpoint strategy to support scalable multi-user experiences without relying on individual calibration.
While this approach simplifies deployment, it does not fully account for variations in user characteristics such as eye height, which may influence depth perception.
Some participants, particularly those with higher eye levels, reported that the extended virtual space felt visually misaligned or compressed under higher $d_v$ conditions. Rather than adapting content to each individual, future systems could explore group-based adaptation, where users are clustered by shared physical traits (\textit{e.g.}, average eye height), and the viewing configuration is optimized at the group level. This strategy may strike a balance between scalability and perceptual accuracy, offering a practical pathway for multi-user XR deployments.

\section{Conclusion and Future Work}

This study explored how a wall display-based XR space can enable shared depth perception by leveraging a 3D scene rendered from a fixed viewpoint rendered from a universal viewpoint. Without relying on HMDs or user-specific tracking, we aimed to support multiple users in experiencing an extended virtual space using innate visual cues. Through a controlled user study, we investigated how variations in virtual depth ($d_v$) and absolute viewing distance ($d_a$) influenced three key perceptual metrics: perceived distance difference (PDD), viewing angle threshold (VAT), and perceived presence. Our findings revealed that as $d_v$ increased, participants significantly underestimated depth, VAT narrowed, and perceived presence—including realness, spatial presence, and visibility—tended to decline. Despite these limitations, participants reported a strong sense of immersion and depth perception within a depth perception zone, even without personalized rendering. Based on these results, we derived three design implications for creating perceptually coherent XR content: (1) adjusting content placement to compensate for depth underestimation, (2) using VAT to define shared depth perception zones, and (3) calibrating the $d_v$ to $d_a$ ratio to optimize presence without perceptual distortion.

While our system is grounded in universal settings to support scalable multi-user access, several future research agendas remain. First, we should explore how group dynamics—such as shared attention, positioning, and conversation—affect depth perception. Future research should also systematically investigate how individual visual cues (\textit{e.g.}, shadows, vanishing points, motion parallax) quantitatively influence the boundaries of this `zone of viewpoint tolerance'. Moreover, comparing these findings against real-world depth perception and other display modalities would fully contextualize the unique perceptual characteristics of wall-sized XR displays. Finally, although we adopted a universal camera height for generalizability, participants' feedback revealed that variations in eye level may influence depth perception. Instead of fully individualizing the system, future XR spaces could employ group-based adaptation, clustering users by physical traits such as eye height to ensure perceptual consistency while maintaining scalability. Based on our findings, future wall display-based XR systems may enable shared, immersive experiences that bridge physical and virtual worlds. Such systems have the potential to support remote cultural exhibition, collaborative environments, and immersive installations, allowing groups of users to access extended virtual spaces together without wearable devices or personalized calibration.

\acknowledgments{%
This research was supported by the Institute of Information \& communications Technology Planning \& Evaluation (IITP) grant funded by the Korea government (MSIT) (No. RS-2024-00397663, Real-time XR Interface Technology Development for Environmental Adaptation, 50\%), National Research Council of Science and Technology (NST) funded by the Ministry of Science and ICT (MSIT), Republic of Korea (No. CRC 21013, 40\%), and by Institute of Information \& communications Technology Planning \& Evaluation (IITP) under the metaverse support program to nurture the best talents (IITP-2025-RS-2022-00156435, 10\%) grant funded by the Korea government (MSIT).
}

\bibliographystyle{abbrv-doi-hyperref}

\bibliography{template}

\end{document}